\documentclass[useAMS,usenatbib]{mn2e}

\usepackage{color}

\usepackage{graphicx}

\newcommand{\rxte}{\emph{RXTE}}

\newcommand{\integral}{\emph{INTEGRAL}}

\newcommand{\wsim}{\ensuremath{\sim}}

\newcommand {\SXL}{Swift~J174510.8$-$2624}
\newcommand {\SX}{Swift~J1745$-$26}
\newcommand{\GX}{GX~339$-$4}
\newcommand{\ST}{XTE~J1752$-$223}
\newcommand{\swift}{\emph{Swift}}

\newcommand {\HS}{H1743$-$322}
\newcommand{\FF}{XTE~J1550$-$564}
\newcommand{\SF}{GRO~J1655$-$40}
\newcommand{\FU}{4U~1543$-$47}

\title{Multiwavelength observations of the black hole transient Swift J1745$-$26 during the outburst decay}

\author[E. Kalemci et al.]{
E. Kalemci$^{1}$\thanks{E-mail:ekalemci@sabanciuniv.edu}, M. \"{O}zbey Arabac\i\ $^{2}$, T. G\"{u}ver$^{3}$, D. M. Russell$^{4,5}$, J. A. Tomsick$^{6}$, 
\newauthor
J. Wilms$^{7}$, G. Weidenspointner$^{8,9}$, E. Kuulkers$^{10}$, M. Falanga$^{11}$, T. Din\c{c}er$^{1}$,
\newauthor
S. Drave$^{12}$, T. Belloni$^{13}$, M. Coriat$^{14}$, F. Lewis$^{15,16}$, T.~Mu\~noz-Darias$^{17}$
\\
$^{1}${Faculty of Engineering and Natural Sciences, Sabanc\i\ University, Orhanl\i-Tuzla, 34956, Istanbul, Turkey}\\
$^{2}${Physics Department, Middle East Technical University, 06530, Ankara, Turkey}\\
$^{3}${Astronomy and Space Sciences Department, Istanbul University, 34119, Universite, Istanbul, Turkey}\\
$^{4}$ New York University Abu Dhabi, P.O. Box 129188, Abu Dhabi, United Arab Emirates\\
$^{5}${Instituto de Astrofisica de Canarias, C/ Via L\'{a}ctea, s/n E38205 - La Laguna (Tenerife), Spain}\\
$^{6}${Space Sciences Laboratory, 7 Gauss Way, University of California, Berkeley, CA, 94720-7450, USA}\\
$^{7}$Dr. Karl-Remeis-Sternwarte and Erlangen Centre for Astroparticle Physics, Friedrich Alexander Universit\"{a}t Erlangen-N\"{u}rnberg, \\
Sternwartstr. 7, 96049 Bamberg, Germany \\
$^{8}$European X-ray Free Electron Laser facility GmbH, Albert-Einstein-Ring 19, 22761, Hamburg, Germany\\
$^{9}${Max Planck Institut f\"{u}r Extraterrestrische Physik, Garching, Germany}\\
$^{10}${European Space Agency, European Space Astronomy Centre, P.O. Box 78, 28691, Villanueva de la Canada, Madrid, Spain}\\
$^{11}${International Space Science Institute (ISSI), Hallerstrasse 6,CH-3012 Bern, Switzerland}\\
$^{12}$School of Physics and Astronomy, Faculty of Physical Sciences and Engineering, University of Southampton, \\
University Road, Southampton, SO17 1BJ, UK\\
$^{13}${INAF - Osservatorio Astronomico di Brera, Via E. Bianchi 46, I-23807, Merate, Italy}\\
$^{14}${Department of Astronomy, University of Cape Town, Private Bag X3, Rondebosch, 7701, South Africa}\\
$^{15}$Faulkes Telescope Project, University of South Wales, Pontypridd, CF37 1DL, Wales\\
$^{16}$Astrophysics Research Institute, Liverpool John Moores University, IC2, Liverpool Science Park, 146 Brownlow Hill, Liverpool L3 5RF, UK\\
$^{17}$University of Oxford, Department of Physics, Astrophysics, Keble Road, Oxford, OX1 3RH, United Kingdom
}

\begin{document}

\date{Accepted for publication by the Monthly Notices of the Royal Astronomical Society}

\pagerange{\pageref{firstpage}--\pageref{lastpage}} \pubyear{2014}

\maketitle


\begin{abstract}

We characterized the broad-band X-ray spectra of \SX\ during the decay of the 2013 outburst using \integral\ ISGRI, JEM-X and \swift\ XRT. The X-ray evolution is compared to the evolution in optical and radio.  We fit the X-ray spectra with phenomenological and Comptonization models. We discuss possible scenarios for the physical origin of a $\wsim$50 day flare observed both in optical and X-rays $\wsim$170 days after the peak of the outburst. We conclude that it is a result of enhanced mass accretion in response to an earlier heating event. We characterized the evolution in the hard X-ray band and showed that for the joint ISGRI-XRT fits, the e-folding energy decreased from 350 keV to 130 keV, while the energy where the exponential cut-off starts increased from 75 keV to 112 keV as the decay progressed. We investigated the claim that high energy cut-offs disappear with the compact jet turning on during outburst decays, and showed that spectra taken with HEXTE on \rxte\ provide insufficient quality to characterize cut-offs during the decay for typical hard X-ray fluxes. Long \integral\ monitoring observations are required to understand the relation between the compact jet formation and hard X-ray behavior. We found that for the entire decay (including the flare), the X-ray spectra are consistent with thermal Comptonization, but a jet synchrotron origin cannot be ruled out.  

\end{abstract}




\section{Introduction}\label{sec:intro}

Galactic black hole transients (GBHT) are excellent laboratories to study the complex relation between the jet and the accretion environment in X-ray binaries as the outbursts evolve on timescales of months, allowing detailed investigation of the properties of accretion states (traced by X-ray spectral and timing properties) together with the properties of jets traced by the radio, and optical/infrared (OIR) emission. 

\begin{figure}
\includegraphics[width=88mm]{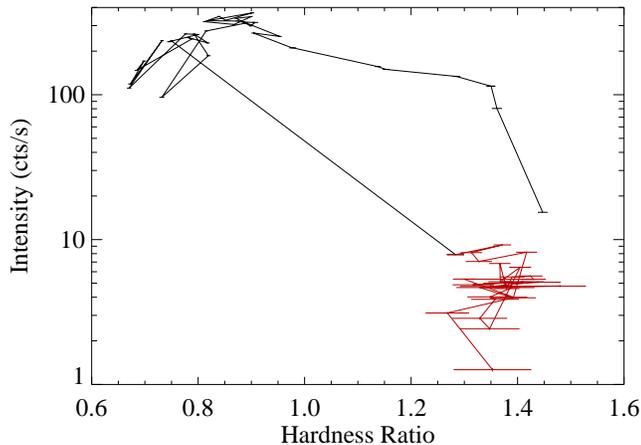}
\caption{\label{fig:hid}
Hardness-Intensity diagram of \SX\ obtained using the \swift\ XRT data only. The hardness ratio is defined as ratio of counts in 2.5$-$10 keV and 1$-$2.5 keV bands and the intensity is the count rate in 1$-$10 keV band. Red data points represent the observations used in this work.
}
\end{figure}

The evolution of accretion states throughout an outburst is best described by the hardness intensity diagram (HID, \citealt{Homan01,Belloni10_jp}). The HID of \SX\ is given in Figure~\ref{fig:hid}. At the start of a typical outburst, the GBHT is in the hard state (HS), the lower right side of the HID.  In this state, the X-ray spectrum is dominated by a hard, power-law like component, historically associated with Compton scattering of soft photons by a hot electron corona. Weak emission from a cool, optically thick, geometrically thin disk (modeled by a multi-temperature blackbody, \citealt{Makishima86}) may also be observed. The variability is strong (typically $>$20\% fractional rms amplitude). As the X-ray flux increases, the GBHT usually makes a transition to the soft state (upper-left side of the HID) in which the X-ray spectrum is dominated by the optically thick accretion disk. At the end of outbursts, sources go back to the hard state as the X-ray flux decreases. There are also intermediate states occurring during transitions between the soft and the hard states (see \citealt{McClintock06book} and \citealt{Belloni10_jp} for details of spectral states of GBHTs). 

In some cases, the evolution of spectral states during an outburst does not follow the usual pattern in the HID. The source can stay in the HS throughout the outburst or make a transition to a hard intermediate state while never fully entering the soft state. These are called "failed outbursts"  \citep{Brocksopp04, Soleri13, Russell13a}. The HID of \SX\ given in Figure~\ref{fig:hid} is a typical HID for a failed outburst \citep[see][for other examples of failed outburst HIDs]{Capitanio09, Ferrigno12, Soleri13}.

\begin{figure*}
\includegraphics[width=150mm]{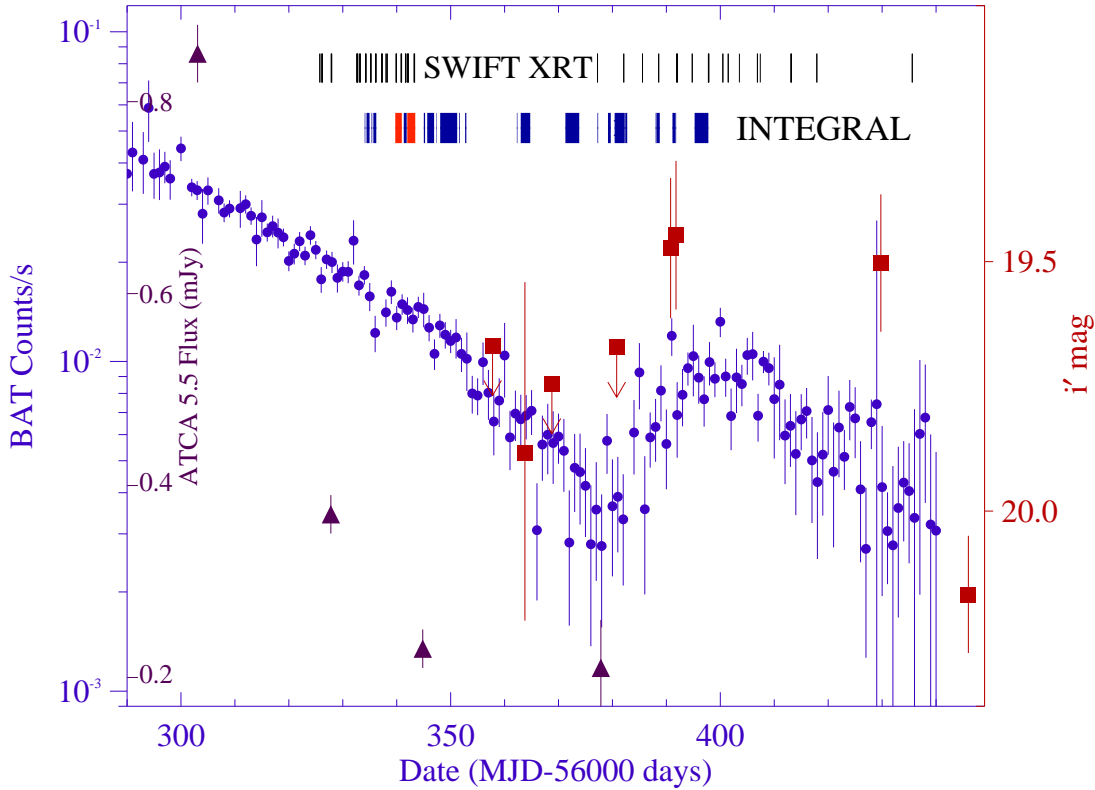}
\caption{\label{fig:allobs}
Multi-wavelength observations used in this work. Blue solid circles represent \swift\ BAT count rate, solid orange squares are i$\prime$ magnitudes (from FTS and RTT 150), purple solid triangles are ATCA radio fluxes at 5.5 GHz. The black short lines at the top represent the times of \swift\ pointed observations. The regions below the \swift\ observations are the times for which the source is within 10$^{\circ}$ of the INTEGRAL pointing direction. The lighter region (red in the coloured version) represents dedicated \integral\ observations (PI Kalemci).}
\end{figure*}

Contemporaneous multi-wavelength observations of GBHTs in optical, near infrared (OIR) and radio show that the behaviour of the jets are closely related to X-ray spectral states. Steady, compact jets are observed in the HS \citep{Corbel00, Fender01b, Fender10_jp, Fender14}.  The typical SEDs in the HS show a flat or inverted power-law at radio frequencies that breaks, at near- to mid-infrared wavelengths, to a second power-law with negative spectral index. Such an emission profile is consistent with emission from a compact, conical jet \citep{Blandford79, Hjellming88}. On the other hand, the radio emission from the jet is quenched in the soft state \citep{Corbel02, Russell11}. During the transition to the soft state in the outburst rise, bright relativistic flares with optically thin radio spectra are sometimes observed \citep[see][and references therein]{Fender14}. 

Two important aspects of the relation between X-ray spectral and timing properties and the jets are understanding the conditions in the accretion environment for the formation of compact jets \citep[][and references therein]{Kalemci13}, and whether jets affect the X-ray spectral and timing properties by altering the accretion environment. One way of achieving these objectives is characterizing the X-ray spectral evolution along with the OIR and radio evolution during outburst decays because the jet emission that is quenched during the soft state turns back on as the source makes a transition to the hard state. 

Such multiwavelength characterization during the outburst decays reveal OIR flares (also called "secondary maxima") when the source is back in the hard state with the X-ray spectrum close to its hardest level \citep{Buxton04, Kalemci05, Coriat09, Russell10, Dincer12, Buxton12, Kalemci13}. In the 2011 decay of \GX, simultaneous radio and OIR observations show that the OIR flare corresponds to a transition in radio from optically thin to optically thick emission \citep{Corbel12, Corbel13}.  Moreover, the spectral energy distributions (SED) created from data during the OIR flares of \FU\ \citep{Buxton04, Kalemci05}, and \FF\ \citep{Jain01_b, Russell10} are consistent with emission from a compact jet. A study by \cite{Kalemci13} showed that the multi-wavelength behavior of most GBHTs that undergo a soft to hard state transition during the outburst decay are similar, such that a compact, steady jet is formed 6-15 days after the initial changes in timing properties, and the jet manifests itself either as an OIR flare, and/or radio emission with flat to inverted spectrum. Alternative explanations for these OIR flares exist, such as the synchrotron emission from a hot accretion flow \citep{Veledina13,Poutanen13}, and irradiation induced secondary mass accretion which could explain the secondary OIR flares with simultaneous X-ray flares \citep{Kalemci13}. 

To understand the effects of the jet on the X-ray spectral properties, we started a program with the \integral\ observatory to characterize the GBHTs at high energies when compact jets are present (or while they form). Thermal Comptonization models that are often invoked to explain the high energy behavior of GBHTs predict  a cut-off in the X-ray spectrum at a few hundred keVs \citep{Gilfanov10}.  While such cut-offs are observed commonly in many GBHTs, in some cases these cut-offs disappear after the jets turn on (such as in \FU, \citealt{Kalemci05}), or are not present at all while the jets are present (\HS\ during 2003 decay, \citealt{Kalemci06}; \GX during the 2005 outburst, \citealt{Kalemci06_mqw}; XTE J1720$-$318, \citealt{CadolleBel04}; \SF\ during the 2005 decay, \citealt{Caballero07}, but also see \citealt{Joinet08}). This disappearance could be interpreted as the jet changing the electron energy distribution and hardening the X-ray spectrum. 

Synchrotron emission from compact jets may also  be contributing to hard X-rays \citep{Markoff01,Markoff05,Maitra09,Russell10}, and both Comptonization and jet models can fit the spectra equally well \citep{Nowak11}. According to \cite{Peer12}, a distinct feature of synchrotron radiation at hard X-rays would be a photon index of around 1.5  before a gradual break in the spectrum at \wsim 10 keV caused by rapid cooling of electrons. Characterizing properties of spectral breaks at hard X-rays therefore is important to test predictions of jet models, especially when these models incorporate detailed electron acceleration mechanisms

Using HEXTE on \rxte, \cite{Miyakawa08} investigated the relation between the cut-off energy (from the fits with cut-off power-law) and several other spectral parameters from all hard state observations of \GX, and found an inverse correlation between the cut-off energy and luminosity for the bright hard state observations. The statistics were not good enough to constrain the evolution of the cut-off parameters for the fainter observations. \cite{Motta09} utilized \integral\ ISGRI and HEXTE together and investigated the evolution of the cut-off energy from the hard state to the  hard intermediate state in the rising phase of the outburst of \GX\ during the 2006/2007 outburst. Similar to findings of \cite{Miyakawa08}, the cut-off energy decreased with increasing flux, but it reversed the trend and increased again with the softening in the hard-intermediate state. While the cause of this behavior is not clear, obtaining the evolution of spectral breaks during the outburst decay, and comparing with the outburst rise could help us to uncover the reason. Since the hard X-ray fluxes during the outburst decays are lower, dedicated observing programs with \integral\ are required for such investigation.

Within our \integral\ program, we first observed \ST\ with \integral, \swift\ and \rxte\ during its 2012 outburst decay \citep{Chun13}. The observation took place a few days before the detection of the compact core with the  Very Long Baseline Array (VLBA). To increase the statistics of the X-ray spectra at high energy we combined all available ISGRI data, and the resulting spectrum required a break in the hard X-ray spectrum. As a continuation of our observing program with \integral, we triggered on \SX\ during its decay in 2013. We also obtained data from \integral\ revolutions before and after our observation.

\begin{table*}
\centering
\begin{minipage}{100mm}
\caption{\integral\ observation details\label{tabobs}}
\begin{tabular}{ccccc}
\hline
ISGRI Rev\footnote{ISGRI Revolution} & Dates & Swift ObsID\footnote{Swift observation within the given revolution} & ISGRI Exposure & Avg. offset\footnote{Average offset of ISGRI pointings} \\
 & MJD-56000 & & (ks) & \\
\hline
1261 & 334.0$-$335.0 & 00032700005 &  61.0 & 3.93$^\circ$ \\
1262 & 335.3$-$336.2 & 00032700007 &  48.0 & 5.19$^\circ$ \\
1263 & 339.7$-$340.9 & 000327000010 &  103.7 & 4.18$^\circ$ \\
1264 & 341.3$-$343.4 & 000327000013 &  151.9 & 4.82$^\circ$ \\
1265 & 345.0$-$346.9 & - &  35.0 & 6.18$^\circ$ \\
1266 & 347.3$-$349.8 & - &  20.8 & 5.31$^\circ$ \\
1267 & 351.0$-$352.9 & - &  29.2 & 5.94$^\circ$ \\
1271 & 362.2$-$364.8 & - &  27.3 & 6.05$^\circ$ \\
1274 & 371.2$-$373.2 & - &  22.7 & 6.17$^\circ$ \\
1276 & 377.2$-$379.7 & 00533836047 &  104.0 & 5.27$^\circ$ \\
1277 & 380.2$-$382.7 & - &  184.2 & 6.34$^\circ$ \\
1279 & 388.0$-$388.8 & 00032700017 &  61.8 & 4.85$^\circ$ \\
1280 & 391.1$-$391.8 & 00032700018 &  44.9 & 5.13$^\circ$ \\
1282 & 395.2$-$397.8 & 00533836049 &  129.5 & 8.50$^\circ$ \\
\hline
\end{tabular}
\end{minipage}
\end{table*}

\subsection{\SX}

\SX\ (\SXL) was discovered by the BAT instrument on board the \swift\ satellite on September 16 2012 (MJD 56186.39, \citealt{Cummings12GCN}). It was subsequently detected by the \swift\ XRT \citep{Sbarufatti12ATEL}, and \integral\ \citep{Vovk12ATEL}. The infrared counterpart was identified by \cite{Rau12}. Based on the X-ray spectral and timing properties, \SX\ is reported to be a Galactic black hole transient \citep{Belloni12ATEL, Tomsick12ATEL}. Further evidence for the nature of the source comes from optical observations that reveal  a broad, double peaked $H_{\alpha}$ emission line whose properties resemble those typically seen in other black hole sources in outburst \citep{UgarteATEL12,MunozDarias13}. 

\SX\ is also detected in radio both with the Karl G. Jansky Very Large Array (VLA, \citealt{MillerJ12ATEL}) and at the Australia Telescope Compact Array (ATCA, \citealt{Corbel12ATEL}). The spectral index of radio observations during the initial hard state is consistent with optically-thick synchrotron emission from a partially self-absorbed compact jet \citep{Corbel12ATEL}. \cite{Curran14} present detailed analysis of radio data from ATCA, VLA and Karoo Array Telescope (KAT-7) monitoring observations of \SX\ until MJD56380. After the initial hard state rise, the source stayed in the hard intermediate state until MJD~56270 based on X-ray spectral and timing properties \citep{Belloni12ATEL,Grebenev12ATEL} and also radio emission properties \citep{Curran14}. Detailed \swift\ and \integral\ analysis of the outburst rise is in preparation (Del Santo et~al.).

Due to the proximity of the source to the Sun, the source could not be observed between MJD~56270 and MJD~56288 with \swift\ BAT, and the first pointed observation of the source with \swift\ XRT after the gap was on MJD~56334. As shown in Figure~1 of \cite{Curran14}, the system was in the hard intermediate state at MJD~56250 as indicated by the inverted spectrum radio emission. The BAT flux shows an increasing trend after this date, indicating that the spectrum remained hard until the start of the Sun gap on MJD~56270. After a gap of 18 days, the BAT light curve before the break smoothly connects to the light curve after the break. The radio spectrum taken on MJD~56304 and the X-ray spectra on  MJD~56334 \citep{Sbarufatti13ATEL} show that the source is in the hard state after the break. This can also be seen in the hardness-intensity diagram given in Figure~\ref{fig:hid}, where the red data points represent observations analyzed in this work. While we cannot rule out the possibility that the source entered the soft state between MJD~56270 and MJD~56304, the trends in the \swift\ BAT light curve and the radio evolution indicate a "failed" outburst. We note that, while the properties of the source are consistent with the definition of failed outbursts, the outburst certainly did not fail in terms of brightness, and had reached close to a Crab during the initial rise \citep{Sbarufatti12ATEL}.

Finally, the source exhibited a secondary flare after MJD~56380 in optical and X-rays \citep{RussellATEL13}. Thanks to the large field of view of \integral\ and the position of the source in the Galactic Bulge, we were able to analyse \integral\ data from several revolutions to characterise the high energy behaviour of the source over 70 days during its decay. In this article, we first present the X-ray evolution of the source using both \integral, and \swift. Then we discuss the secondary flare observed in both optical and X-ray and compare this flare with the secondary OIR flares observed in GBHTs.


\section{Observations and Analysis}\label{sec:obs}

\subsection{X-ray observations}

A summary plot of all multi-wavelength observations during the decay can be found in Figure~\ref{fig:allobs}. The source has been covered very well with \swift\ and \integral. Our analysis includes \swift\ data between MJD~ 56334.23 and MJD~56435.59, and \integral\ data between MJD~56334.0 and MJD~56397.8. The details of the \integral\ revolutions we utilized and simultaneous \swift\  ObsIDs are given in Table~\ref{tabobs}.

\subsection{Spectral extraction with \swift\ XRT, \integral\ ISGRI and JEM-X}\label{subsec:specxrt}

The \swift\ XRT observations used in this study are shown in Table~\ref{xrtonly}. All data were processed following the standard procedures and using the {\it xrtpipeline v.0.12.8}. Since the source flux varies significantly, the whole sample of pointings contain both Windowed Timing (WT) mode and Photon Counting (PC) mode data. In order to minimise the affects of photon pile-up we followed the method outlined in \cite{Reynolds13} and created source selection regions based on the observed count rate. The background region was selected from an annulus with an inner and outer radius of 70$"$ and 100$"$, respectively, from the source centroid. Events were selected with grades 0-2 and 0-12 for WT and PC mode data, respectively. The appropriate response matrix files,  swxwt0to2s6\_20010101v014.rmf for the WT mode and swxpc0to12s6\_20010101v013.rmf for the PC mode,   were obtained from HEASARC CALDB. The auxiliary response matrix files were created using the HEASOFT tool {\it xrtmkarf} and the exposure map created by the {\it xrtexpomap}.   Finally we grouped all the spectra to have at least 50 counts per spectral bin. After the pile-up mitigating procedure, the PC mode observations resulted in a spectrum with few energy bins and large errors, and therefore are not used in this work.

The \emph{INTEGRAL} data were reduced and analyzed using the standard Off-line Scientific Analysis (OSA) software package (v.10) released by the ISDC \citep{Courvoisier03} for each revolution. The \integral\ revolutions last  $\wsim$3 days, and within revolutions sky regions are observed with 30-60 minute individual pointings with a special dither pattern to minimise noise in mosaic images obtained by combining images from individual pointings.

For JEM-X, we combined images in a single mosaic per revolution in order to reach the highest sensitivity for JEM X-1 and JEM X-2 separately. The spectra were extracted from the each combined mosaic image over an energy range of 3-35 keV corresponding to 16 channels. After comparing the quality of spectra, we have decided to use only JEM X-1 data. In the case of IBIS/ISGRI, the standard procedure was used to extract the spectra and rebin the response matrix to 50 bins. Then we used \emph{spe\_pick} tool to obtain average spectrum of the source for each revolution for 18-350 keV energy range.

\subsection{X-ray spectral analysis}\label{subsec:spec}

We first fitted \swift\ XRT and \integral\ ISGRI spectra separately. For the \swift\ XRT fits, we started with interstellar absorption ($tbabs$ in XSPEC) and a power-law. We used cross sections of \cite{Verner96} and abundances of \cite{Wilms00} for the interstellar absorption and left $N_{H}$ as a free parameter initially. We observed significant residuals at energies less than 2 keV. We added a multicolor disk blackbody ($diskbb$ in XSPEC, \citealt{Makishima86}) which resulted in fits with acceptable reduced $\chi^{2}$ values. The results of the fits are tabulated in Table~\ref{xrtonly}\footnote{All errors in the figures and in tables correspond to $\Delta\chi^{2}$ of 2.706.}

\begin{figure}
\includegraphics[width=88mm]{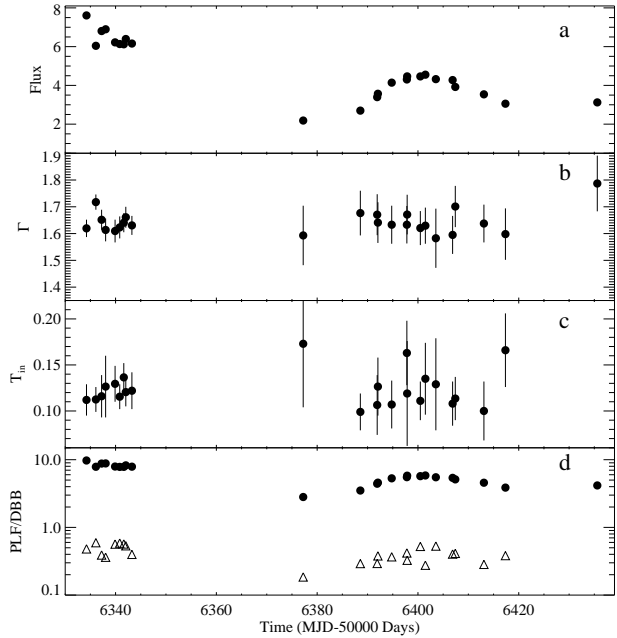}
\caption{\label{fig:onlyxrtv2}
Plot of X-ray spectral parameters of \swift\ XRT data only, fitted with a power-law+diskbb model, interstellar absorption (tbabs) with $N_{H}$ fixed to $2.18 \times\ 10^{22} cm^{-2}$. a: 1-12 keV flux in units of $10^{-10}$ ergs cm$^{-2}$ s$^{-1}$. b: power-law photon index, c: diskbb inner disk temperature in keV, d:  PLF (solid circle) is the power law flux and DBB (triangle) is the diskbb flux in units of  $10^{-10}$ ergs cm$^{-2}$ s$^{-1}$ in 1-12 keV band.
}
\end{figure}

For the initial, high flux, high quality spectra between MJD~55633.4 and MJD~55644.3, we realized that fits result in a small scatter of $N_{H}$ values around an average of $2.18 \times 10^{22}$ atoms cm$^{-2}$. Lower quality spectra showed large scatter in $N_{H}$ that correlated with disk fit parameters. Since it is unlikely that $N_{H}$ varies more than 20\% within timescale of days, we fixed its value to $2.18 \times 10^{22}$ atoms cm$^{-2}$ and performed the fits again. The results show a smooth evolution of X-ray spectral parameters as seen in  Figure~\ref{fig:onlyxrtv2} and in Table~\ref{xrtonly}.

\begin{table*}
\begin{minipage}{160mm}
\centering
\caption{Swift XRT only fit results\label{xrtonly}}
\begin{tabular}{cc|ccc|cccc}
\hline \hline
Swift obsid & Date\footnote{MJD-56000} &\multicolumn{3}{c}{Fits with free $N_{H}$} & \multicolumn{4}{c}{Fits with $N_{H}$ fixed} \\
 & & $N_{H}$\footnote{in units of $10^{22}$ atoms cm$^{-2}$} & $\Gamma$ & $T_{in}$ & $\Gamma$ & $T_{in}$ & Flux\footnote{1-12 keV Flux in units of $10^{-10}$ ergs cm$^{-2}$ s$^{-1}$} & Norm\footnote{Lower limit of normalization of \emph{diskbb} component}\\
\hline
00032700005 & 334.2 & 2.12$\pm$0.23 & 1.60$\pm$0.07 & 0.10$\pm$0.02 & 1.62$\pm$0.03 & 0.11$\pm$0.02 & 7.61$\pm$0.10& $>$1.07E+06\\
00032700007 & 336.1 & 2.08$\pm$0.24 & 1.68$\pm$0.08 & 0.11$\pm$0.02 & 1.72$\pm$0.03 & 0.11$\pm$0.01 & 6.04$\pm$0.09& $>$1.64E+06\\
00032700008 & 337.2 & 2.14$\pm$0.27 & 1.64$\pm$0.08 & 0.11$\pm$0.03 & 1.65$\pm$0.04 & 0.12$\pm$0.02 & 6.81$\pm$0.08& $>$2.41E+05\\
00032700009 & 338.0 & 1.88$\pm$0.15 & 1.53$\pm$0.06 & 0.07$\pm$0.03 & 1.61$\pm$0.04 & 0.13$\pm$0.03 & 6.90$\pm$0.08& $>$1.15E+05\\
00032700010 & 339.9 & 2.08$\pm$0.30 & 1.58$\pm$0.08 & 0.12$\pm$0.03 & 1.61$\pm$0.04 & 0.13$\pm$0.02 & 6.22$\pm$0.09& $>$3.00E+05\\
00032700011 & 341.6 & 2.44$\pm$0.27 & 1.70$\pm$0.07 & 0.14$\pm$0.01 & 1.64$\pm$0.04 & 0.14$\pm$0.02 & 6.12$\pm$0.07& $>$2.52E+05\\
00032700012 & 340.8 & 2.41$\pm$0.28 & 1.69$\pm$0.09 & 0.12$\pm$0.01 & 1.62$\pm$0.04 & 0.12$\pm$0.01 & 6.14$\pm$0.09& $>$1.27E+06\\
00032700013 & 342.0 & 2.26$\pm$0.26 & 1.68$\pm$0.08 & 0.12$\pm$0.02 & 1.66$\pm$0.04 & 0.12$\pm$0.02 & 6.39$\pm$0.09& $>$6.83E+05\\
00032700014 & 343.2 & 2.22$\pm$0.27 & 1.64$\pm$0.07 & 0.12$\pm$0.02 & 1.63$\pm$0.04 & 0.12$\pm$0.02 & 6.16$\pm$0.06& $>$3.52E+05\\
00533686047 & 377.2 & 1.64$\pm$0.23 & 1.49$\pm$0.13 & - & 1.59$\pm$0.11 & 0.17$\pm$0.07 & 2.19$\pm$0.07& $>$2.10E+04\\
00032700017 & 388.6 & 2.76$\pm$0.55 & 1.85$\pm$0.19 & 0.10$\pm$0.02 & 1.68$\pm$0.08 & 0.10$\pm$0.02 & 2.70$\pm$0.09& $>$1.59E+06\\
00032700018 & 391.9 & 2.29$\pm$0.52 & 1.69$\pm$0.17 & 0.10$\pm$0.04 & 1.67$\pm$0.08 & 0.11$\pm$0.03 & 3.39$\pm$0.11& $>$3.14E+05\\
00533836048 & 392.1 & 2.36$\pm$0.55 & 1.68$\pm$0.16 & 0.12$\pm$0.03 & 1.64$\pm$0.08 & 0.13$\pm$0.03 & 3.57$\pm$0.09& $>$1.18E+05\\
00032700019 & 394.8 & 2.21$\pm$0.46 & 1.63$\pm$0.15 & 0.10$\pm$0.03 & 1.63$\pm$0.07 & 0.11$\pm$0.03 & 4.14$\pm$0.10& $>$6.08E+05\\
00533836049 & 397.8 & 3.28$\pm$0.66 & 1.86$\pm$0.15 & 0.15$\pm$0.02 & 1.63$\pm$0.07 & 0.16$\pm$0.03 & 4.31$\pm$0.12& $>$2.10E+04\\
00032700020 & 397.9 & 1.92$\pm$0.33 & 1.59$\pm$0.13 & 0.08$\pm$0.06 & 1.67$\pm$0.07 & 0.12$\pm$0.06 & 4.47$\pm$0.12& $>$4.50E+05\\
00032700021 & 400.5 & 2.66$\pm$0.49 & 1.74$\pm$0.14 & 0.11$\pm$0.02 & 1.62$\pm$0.06 & 0.11$\pm$0.02 & 4.46$\pm$0.10& $>$9.16E+05\\
00533836050 & 401.5 & 2.74$\pm$0.55 & 1.76$\pm$0.15 & 0.14$\pm$0.03 & 1.63$\pm$0.07 & 0.13$\pm$0.04 & 4.55$\pm$0.11& $>$4.06E+05\\
00032700022 & 403.6 & 2.37$\pm$0.88 & 1.62$\pm$0.25 & 0.11$\pm$0.07 & 1.58$\pm$0.11 & 0.13$\pm$0.05 & 4.32$\pm$0.19& $>$5.39E+05\\
00032700023 & 406.9 & 2.84$\pm$0.53 & 1.77$\pm$0.16 & 0.11$\pm$0.02 & 1.60$\pm$0.07 & 0.11$\pm$0.02 & 4.28$\pm$0.12& $>$6.76E+05\\
00533836051 & 407.4 & 2.35$\pm$0.49 & 1.75$\pm$0.16 & 0.11$\pm$0.02 & 1.70$\pm$0.08 & 0.11$\pm$0.02 & 3.92$\pm$0.11& $>$5.14E+05\\
00533836052 & 413.1 & 2.23$\pm$0.47 & 1.64$\pm$0.16 & 0.09$\pm$0.04 & 1.64$\pm$0.07 & 0.10$\pm$0.03 & 3.54$\pm$0.09& $>$5.00E+05\\
00533836053 & 417.4 & 2.20$\pm$0.00 & 1.60$\pm$0.10 & 0.17$\pm$0.04 & 1.60$\pm$0.10 & 0.17$\pm$0.04 & 3.06$\pm$0.10& $>$1.43E+05\\
00533836057 & 435.6 & 3.14$\pm$1.08 & 2.01$\pm$0.32 & 0.08$\pm$0.06 & 1.79$\pm$0.10 & - & 3.12$\pm$0.13& - \\
\end{tabular}
\end{minipage}
\end{table*}

For the \integral\ observations, we first fitted the ISGRI data from individual revolutions. We started with a power-law, and added a high energy cut off component ($highecut$ in XSPEC) and test if addition of the cutoff significantly improves the $\chi^{2}$. If an \emph{F-test} results in chance probability less than 0.001, we used the fit with the cutoff. The fit results are summarized in Table~\ref{table:isgr} and some of the fit parameters are plotted in Figure~\ref{fig:joint}.

Finally, we applied joint fits to the \swift\ XRT, \integral\ ISGRI and JEM-X for observations with concurrent data (Table~\ref{tabobs}). We started with a model that consists of power-law and multicolor diskblackbody components with interstellar absorption. Similar to ISGRI-only fits, we add a high energy cut-off and check the improvement in $\chi^{2}$. If an \emph{F-test} results in chance probability less than 0.001, we used the fit with the cutoff. 

We also applied thermal and hybrid Comptonization models to the joint spectra. In the \emph{compps} model \citep{Poutanen96}, blackbody photons are Compton up-scattered from a spherical corona with uniform optical depth $\tau_{y}$ and electron temperature $kT_{e}$. In our fits, the blackbody temperature is fixed to the \emph{diskbb} inner temperature $T_{in}$. We kept all other parameters at default values. The fits cannot constrain the reflection fraction. 

Finally, we tried the \emph{eqpair} model \citep{Coppi99} to be able to assess the presence of non-thermal electron population. The \emph{eqpair} model assumes a spherical plasma of electrons, positrons and soft photons. The resulting emission is determined by the compactness parameter $l = L \sigma{T}/Rm_{e}c^{3}$, where $L$ is power, $R$ is the radius of the spherical plasma, and $\sigma_{T}$ is the Thomson cross section. The model allows hybrid electron plasmas, thermal and non-thermal electron energy distributions together. Following \cite{Coppi99},  we fixed the soft photon compactness $l_{s}$ to 1, and fit our data with a minimum set of free parameters. The temperature of seed photons are fixed to the inner disk temperature of the \emph{diskbb} component. The fit results for Comptonization models are summarized in Table~\ref{table:joint}. An example fit is shown in Figure~\ref{fig:1263fits}.

\begin{table}
\centering
\begin{minipage}{88mm}
\caption{Optical Magnitudes\label{table:mags}}
\begin{tabular}{cccc}
\hline
Date  & $i'$ & $R$ & $V$  \\ \hline
56357.8 & $>$19.67 & $>$20.75 & $>$19.73\\
56363.8 & 19.88$\pm$0.34 & 20.92$\pm$0.35 & $>$21.13\\
56368.8 & $>$19.74 & 20.67$\pm$0.24 & $>$21.19\\
56380.8 & $>$19.67 & $>$20.83 & $>$20.39\\
56390.8 & 19.47$\pm$0.14 & 20.72$\pm$0.25 & $>$20.83\\
56391.8 & 19.45$\pm$0.15 & 20.45$\pm$0.21 & 21.58$\pm$0.48\\
56429.8 & 19.50$\pm$0.14 & 20.20$\pm$0.13 & 22.94$\pm$0.55\\
56441.8 & $>$18.67 & $>$20.36 & 21.04$\pm$0.55\\
56451.8 & $>$18.98 & $>$20.49 & $>$21.97\\
56537.4 & $>$20.78 & - & $>$21.26\\
56446\footnote{From \cite{Grebenev14}}   & 20.07$\pm$0.12 & - & - \\
\hline
\end{tabular}
\end{minipage}
\end{table}

\subsection{Radio and optical data}

During the decay, the source was also monitored in radio and in optical. There were four radio observations with the ATCA, and for all those observations the radio spectra are consistent with being flat to inverted, indicating presence of compact jets \citep{Curran14}. 

Optical observations were acquired using the Faulkes Telescope South (FTS; located at Siding Spring, Australia) as part of a monitoring campaign of $\wsim 30$ LMXBs with the Faulkes Telescopes \citep{Lewis08}. Three filters were used, Sloan Digital Sky Survey (SDSS) i$^{\prime}$, Bessell R and Bessell V. Bias subtraction and flat-fielding were performed via the automatic pipelines. For flux calibration procedures, see \cite{MunozDarias13}. The magnitudes in each band are given in Table~\ref{table:mags}. We also take one i$^{\prime}$ band magnitude measured on MJD~56446 taken at Turkish National Observatory with Russian Turkish Telescope RTT150 reported in \cite{Grebenev14}.


\section{Results}\label{sec:results}

\subsection{Soft X-ray evolution with \swift\ XRT}

As mentioned earlier, leaving $N_{H}$ free results in highly scattered, and correlated $N_{H}$ and \emph{diskbb} parameters. In some cases the \emph{diskbb} flux changes by an order of magnitude within a day. After fixing the $N_{H}$, a smooth evolution of X-ray spectral fit parameters is observed (Figure~\ref{fig:onlyxrtv2}). 

A soft \emph{diskbb} component is required statistically in the fits, however, its parameters are not very well determined, especially the normalization. The 90\% confidence lower limits on the \emph{diskbb} normalisations are given in Table~\ref{xrtonly} along with other important fit parameters. The minimum normalizations typically are larger than a few times $10^{5}$, and the lowest we measure is $10^{4}$. While the quality of the data is not good enough to place strong constraints on the inner disk radius, for typical black hole masses and distances, these minimum normalization values indicate large inner disk radii.

The 1-12 keV X-ray flux decreases from $8 \times 10^{-10}$ ergs cm$^{-2}$ s$^{-1}$ to $2 \times 10^{-10}$ ergs cm$^{-2}$ s$^{-1}$ in $\wsim$43 days before showing a flare. The flare lasts for approximately 40 days. As seen in Figure~\ref{fig:allobs}, the flare occurs in both soft and hard X-rays. The flare causes an increase of flux in both \emph{diskbb} and power-law flux, and the photon indices remain remarkably constant even though the soft X-ray flux more than doubles during the flare. 

\begin{table*}
\centering
\begin{minipage}{100mm}
\caption{ISGRI only fit results\label{table:isgr}}
\begin{tabular}{ccccc}
\hline
ISGRI Rev  & $\Gamma$ & $E_{fold}$\footnote{Folding energy in \emph{highecut} model} & $E_{Cut}$\footnote{Cut-off energy in \emph{highecut} model} & Flux\footnote{13-200 keV Flux in units of $10^{-9}$ ergs cm$^{-2}$ s$^{-1}$} \\
 & & (keV) & (keV) & \\  \hline
1261 & 1.55$\pm$0.06 & 273.6$\pm$ 94.6 &  68.9$\pm$ 12.6 & 2.82$\pm$0.07\\
1262 & 1.45$\pm$0.20 & $<$110.5  &   $>$178.6  & 2.63$\pm$0.09\\
1263 & 1.56$\pm$0.06 & 255.9$\pm$ 82.3 &  65.7$\pm$ 16.7 & 2.37$\pm$0.04\\
1264 & 1.60$\pm$0.04 & 220.3$\pm$ 62.0 &  81.2$\pm$ 13.4 & 2.30$\pm$0.03\\
1265 & 1.61$\pm$0.07 & 33.0$^{+109.1}_{-21.3}$ & 140.8$^{+13.2}_{-31.9}$  & 2.03$\pm$0.11 \\
1266 & 1.70$\pm$0.09 & - & - & 2.08$\pm$0.07\\
1267 & 1.74$\pm$0.09 & - & - & 1.65$\pm$0.06\\
1271 & 1.62$\pm$0.14 & - & - & 1.11$\pm$0.09\\
1274 & 1.72$\pm$0.23 & - & - & 0.74$\pm$0.24\\
1276 & 1.74$\pm$0.14 & - & - & 0.71$\pm$0.05\\
1277 & 1.84$\pm$0.10 & - & - & 0.76$\pm$0.03\\
1279 & 1.72$\pm$0.10 & - & - & 0.97$\pm$0.04\\
1280 & 1.78$\pm$0.09 & - & - & 1.22$\pm$0.04\\
1282 & 1.64$\pm$0.05 & 131.8$\pm$ 71.5 & 114.2$\pm$ 15.4 & 1.83$\pm$0.05\\
\hline
\end{tabular}
\end{minipage}
\end{table*}

\subsection{Hard X-ray evolution and joint fits}

The evolution of the hard X-ray flux with BAT is shown in Figure~\ref{fig:allobs}, and the evolution of the spectral fit parameters of \integral\ ISGRI data only are given in Table~\ref{table:isgr}. We were able to search and detect cut-offs in the ISGRI data for revolutions with high flux (1261 to 1265), and also during the final revolution (1282) for which the source was in the field of view of \integral. For revolution 1282,  the statistical quality of the spectrum is better than for revolutions 1266 to 1280 because of higher exposure as well as higher flux due to the observation occurring near the peak of the X-ray flare.

\begin{figure}
\includegraphics[width=88mm]{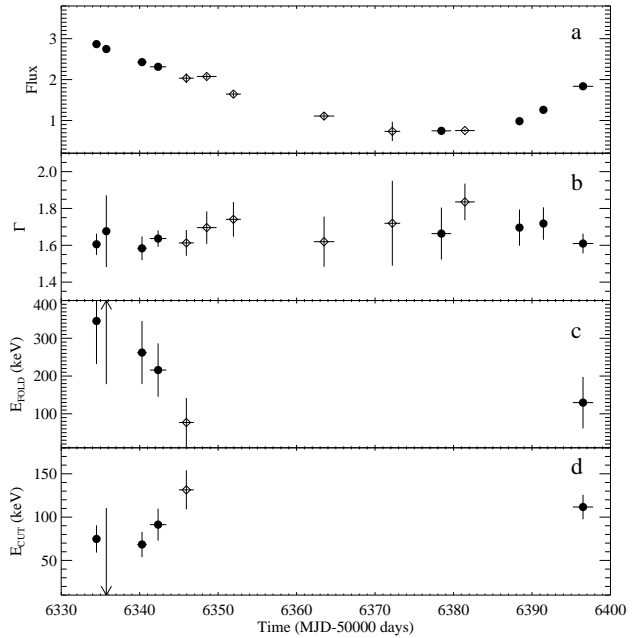}
\caption{\label{fig:joint}
Evolution of spectral fit parameters from a with a power-law and a cut-off. Filled circles are joint fits to \integral\ ISGRI and \swift\ XRT (therefore the model also includes \emph{diskbb}), and empty diamonds are fits to only ISGRI spectra. a: 13-200 keV flux in units of 10$^{-9}$ ergs cm$^{-2}$ s$^{-1}$. b: photon index. c: Folding Energy and d: Cut-off energy of the \emph{highec} model, respectively. 
}
\end{figure}

For \integral\ revolutions with simultaneous \swift\ XRT data we applied joint fits with three models; phenomenological \emph{highec*(power+diskbb)}, and Comptonization models \emph{compps+diskbb} and \emph{eqpair+diskbb}. For revolutions 1261, 1263, 1264 and 1282, a cut-off is required in the power-law model fit (see Table~\ref{table:joint}). The evolution of spectral parameters related to hard X-rays is summarized in Fig.~\ref{fig:joint}. In this figure we used solid circles to represent joint fit results, and used diamonds from ISGRI only fits for which we do not have XRT data.  

At first instance, fits from the earliest five revolutions seem to have a harder photon index compared to later revolutions, however this is an artificial result due to the presence of a cut-off in the spectra of earlier revolutions. This evolution indicates that it is likely that the data from revolutions 1266 to 1280 also require a high energy cut-off, however since it is not required by the \emph{F-test} due to low statistics, the cut-off is not included in the fits and the resulting power-law index is softer.

The evolution of folding and cut-off energies indicate that as the flux decayed initially, the cut-off energies increased and folding energies decreased. The high cut-off, low folding energy behavior was also observed during the peak of the secondary flare during revolution 1282. 

For the \emph{compps} fits, we first let electron temperature, $y-parameter$ ($\tau_{y}$), and reflection fraction free. For all observations, the reflection fraction is close to zero and could not be constrained by the fits. We then fixed this parameter to zero, and fitted the spectra with electron temperature and $\tau_{y}$ as free parameters (see Table~\ref{table:joint} for results). We find electron temperatures of 60-100 keV for the revolutions 1261 to 1264. The electron temperature at revolution 1282 is higher, but with large errors. We cannot constrain the $\tau_{y}$ with \emph{compps} fits, within $\Delta \chi^{2}$ of 2.706, the $\tau_{y}$ parameter is pegged to the upper limit of 3. 

\begin{figure}
\includegraphics[width=88mm]{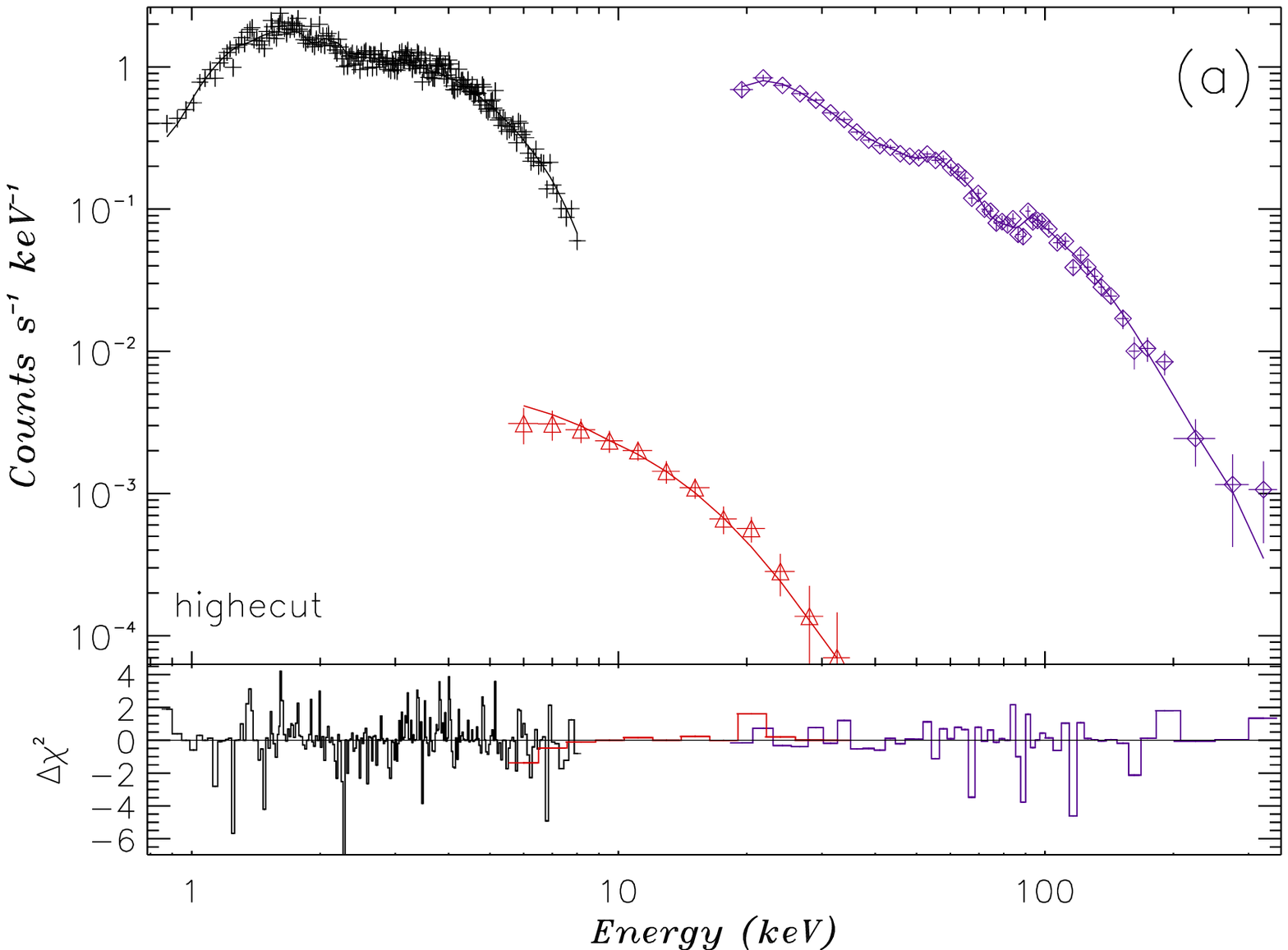}
\includegraphics[width=88mm]{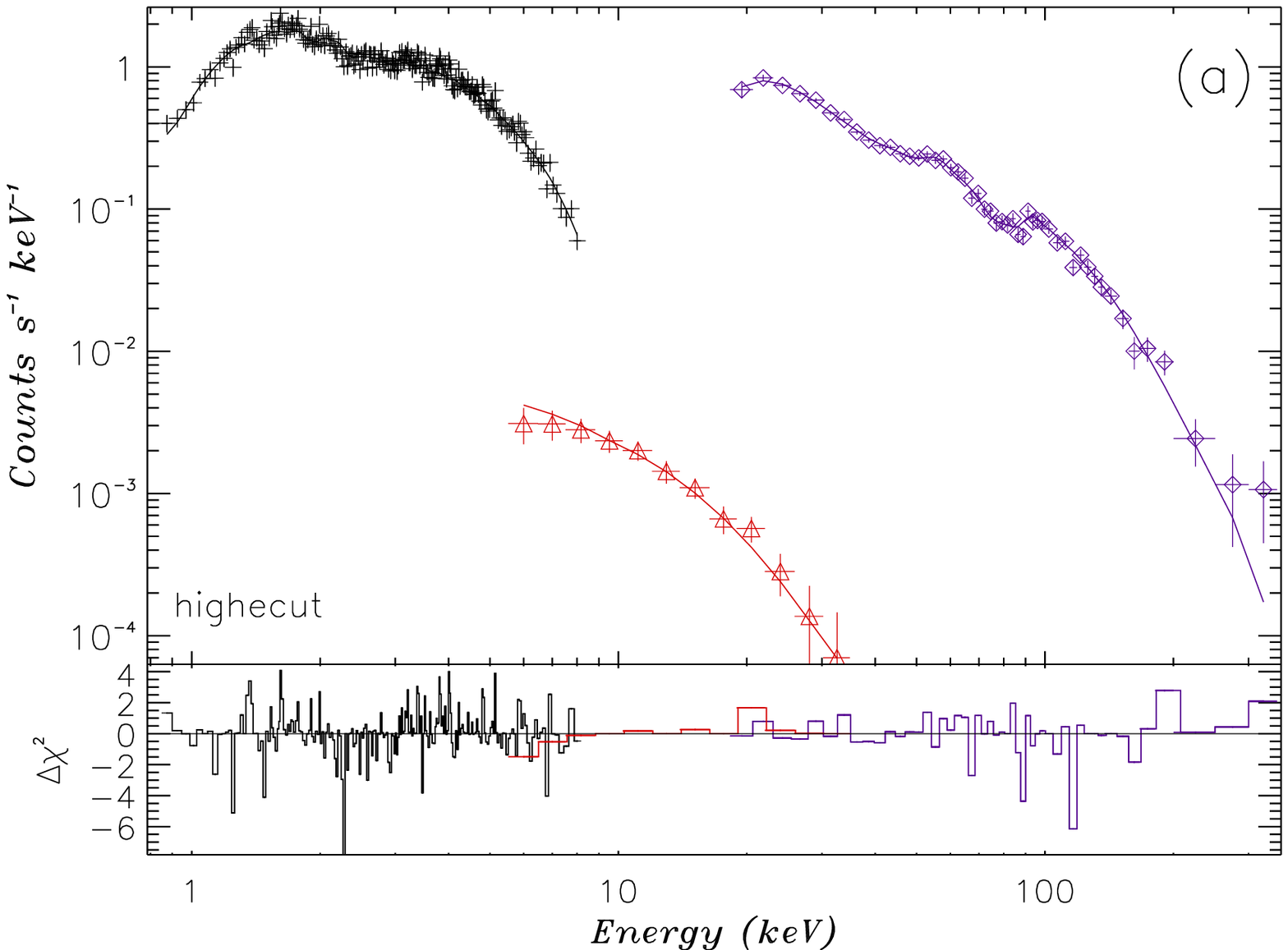}
\includegraphics[width=88mm]{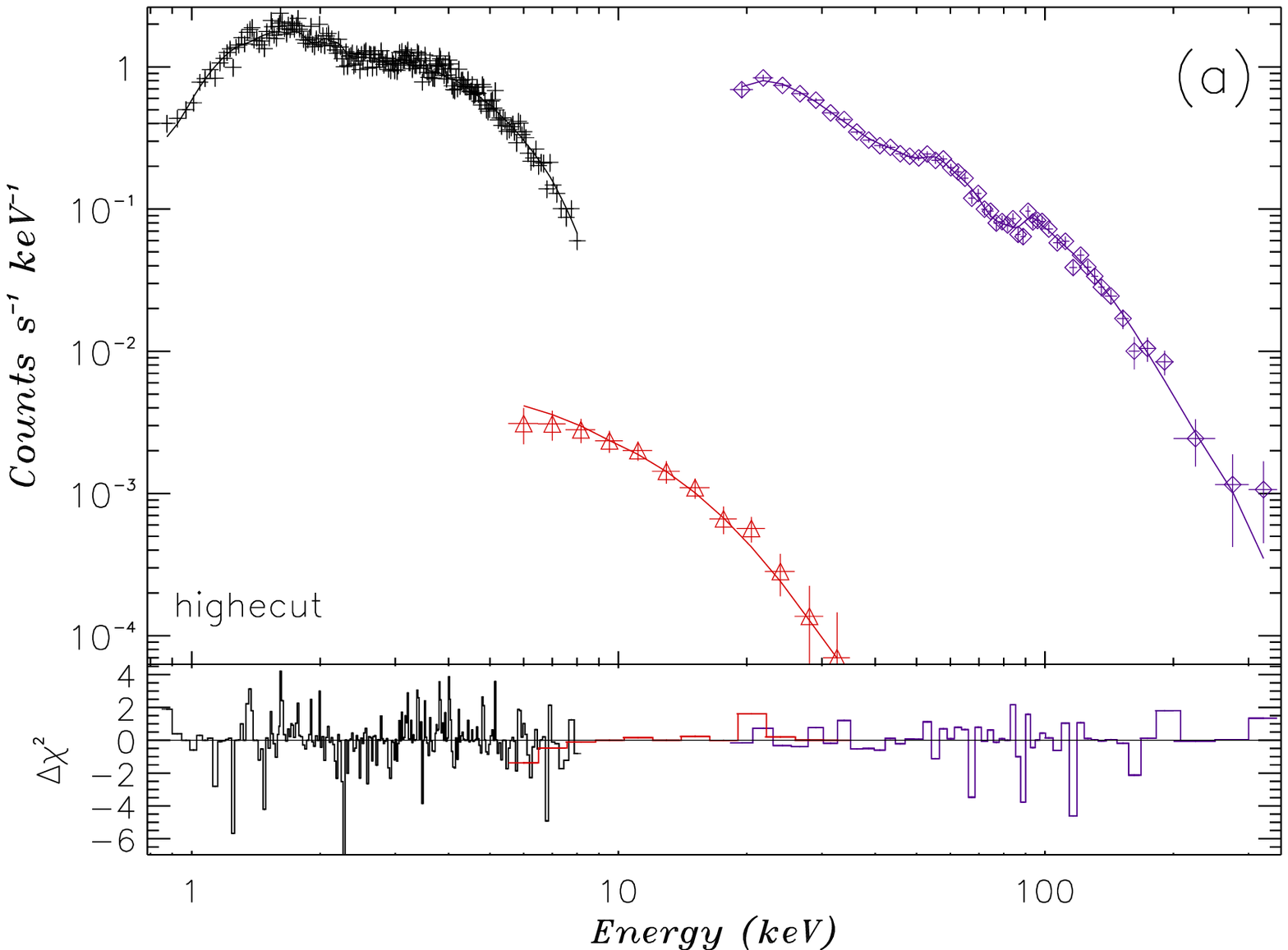}
\caption{\label{fig:1263fits} Fits to joint \swift\ XRT (crosses, black in colored version), \integral\ ISGRI (diamonds, purple in colored version) and JEM-X (triangles, red in colored version) spectra for revolution 1263. Fits are with a: \emph{highecut}, b: \emph{comps} and c: \emph{eqpair} model.
}
\end{figure} 

For the \emph{eqpair} fits we started with fixing $l_{nt}/l_{h}$ to zero to see if the thermal electron distribution can fit the data. We only had $l_{h}/l_{s}$ (hard to soft compactness ratio) and $\tau_{p}$ free and fixed all the other parameters. After we obtained initial fits, we let reflection fraction vary. Like the \emph{compps} case, for all revolutions, the fits resulted in reflection fractions close to zero. We fixed it to zero and repeated the fits. The results are given in  Table~\ref{table:joint}. 

We then tested whether a hybrid electron energy distribution provides better fit to the data. We made $l_{nt}/l_{h}$ a free parameter and fixed $\Gamma_{inj}$ to 2.5. While the fits cannot constrain the $l_{nt}/l_{h}$ parameter well enough to rule out non-thermal plasma, we consistently obtain low values of $l_{nt}/l_{h}$ for all revolutions, except revolution 1262. The 2$\sigma$ limits for this parameter are given in Table~\ref{table:joint}. Except revolution 1262 $l_{nt}/l_{h}$ values are consistent with zero, and the $l_{h}/l_{s}$ and $\tau_{p}$ parameters for the non-thermal fits are very similar to the thermal fits. Therefore, in Table~\ref{table:joint} we only show thermal fit results for these parameters. 

\begin{table*}
\centering
\begin{minipage}{120mm}
\caption{joint fit results\label{table:joint}}
\begin{tabular}{cccccc}
\hline
\multicolumn{6}{c}{High Energy Cut-off Fits}\\
ISGRI Rev  & $\Gamma$ & $E_{Fold}$ & $E_{Cut}$ & Flux\footnote{1-200 keV Flux in units of $10^{-9}$ ergs cm$^{-2}$ s$^{-1}$} & $\chi^{2}$/DOF \\ \hline
1261 & 1.61$\pm$0.02 & 345.8$\pm$113.9 &  74.8$\pm$ 15.7 & 3.69$\pm$0.05 & 407.4/357\\
1262 & 1.68$\pm$0.03 & - & - & 3.422$\pm$0.05 & 310.4/284\\
1263 & 1.58$\pm$0.01 & 262.2$\pm$ 83.5 &  68.4$\pm$ 14.5 & 3.10$\pm$0.07 & 235.0/269\\
1264 & 1.64$\pm$0.03 & 215.8$\pm$ 70.8 &  91.3$\pm$ 18.4 & 3.00$\pm$0.04 & 295.5/275\\
1276 & 1.66$\pm$0.03 & - & - & 0.98$\pm$0.05 &  96.4/ 88\\
1279 & 1.70$\pm$0.06 & - & - & 1.26$\pm$0.05 &  89.0/ 94\\
1280 & 1.72$\pm$0.06 & - & - & 1.61$\pm$0.04 & 122.0/111\\
1282 & 1.61$\pm$0.04 & 129.6$\pm$ 68.2 & 111.6$\pm$ 14.1 & 2.30$\pm$0.05 & 179.7/135\\
\hline
\multicolumn{6}{c}{Compps Fits} \\
ISGRI Rev  & \multicolumn{2}{c}{$kT_{e}$ (keV)} & \multicolumn{2}{c}{$\tau_{y}$} & $\chi^{2}$/DOF \\ \hline
1261 & \multicolumn{2}{c}{ 73.5$\pm$13.4} & \multicolumn{2}{c}{$3^{+0}_{-0.93}$} & 409.0/358\\ 
1262 & \multicolumn{2}{c}{ 92.9$\pm$23.9} & \multicolumn{2}{c}{$3^{+0}_{-1.60}$} & 317.8/283\\ 
1263 & \multicolumn{2}{c}{ 66.3$\pm$ 8.0} & \multicolumn{2}{c}{$3^{+0}_{-0.62}$} & 239.7/270\\ 
1264 & \multicolumn{2}{c}{ 69.3$\pm$11.8} & \multicolumn{2}{c}{$3^{+0}_{-0.89}$} & 295.7/276\\ 
1282 & \multicolumn{2}{c}{119.3$\pm$60.9} & \multicolumn{2}{c}{$3^{+0}_{-2.16}$} & 187.2/136\\ 
\hline
\multicolumn{6}{c}{Eqpair Fits\footnote{$l_{bb}$ is fixed to 1., $\Gamma_{inj}$ is fixed to 2.5 and $Refl$ fixed to 0.}} \\
ISGRI Rev  & $\tau_{p}$ & $l_{h}/l_{s}$ & $\chi^{2}_{ther}$/DOF\footnote{$\chi^{2}$ of thermal electron energy distribution fits, $l_{nt}/l_{h}$=0.} &$l_{nt}/l_{h}$\footnote{2$\sigma$ limits} & $\chi^{2}_{hybr}$/DOF\footnote{$\chi^{2}$ of hybrid electron energy distribution fits, $l_{nt}/l_{h}$ is a free parameter} \\ \hline
1261 &  2.06$\pm$ 0.11 & 12.25$\pm$1.08 & 406.8/357 & $<$0.48 & 406.6/356\\
1262 &  1.23$\pm$ 0.24 &  9.99$\pm$1.12 & 313.1/282 & $>$0.39 & 306.5/281\\
1263 &  2.03$\pm$ 0.31 & 11.51$\pm$1.10 & 240.5/269 & $<$0.50 & 240.5/268\\
1264 &  1.79$\pm$ 0.32 & 10.37$\pm$0.93 & 294.2/275 & $<$0.78 & 293.5/274\\
1282 &  1.32$\pm$ 0.70 &  8.29$\pm$1.18 & 185.4/135 & $<$0.57 & 188.8/134\\
\end{tabular}
\end{minipage}
\end{table*}

\begin{figure}
\includegraphics[width=88mm]{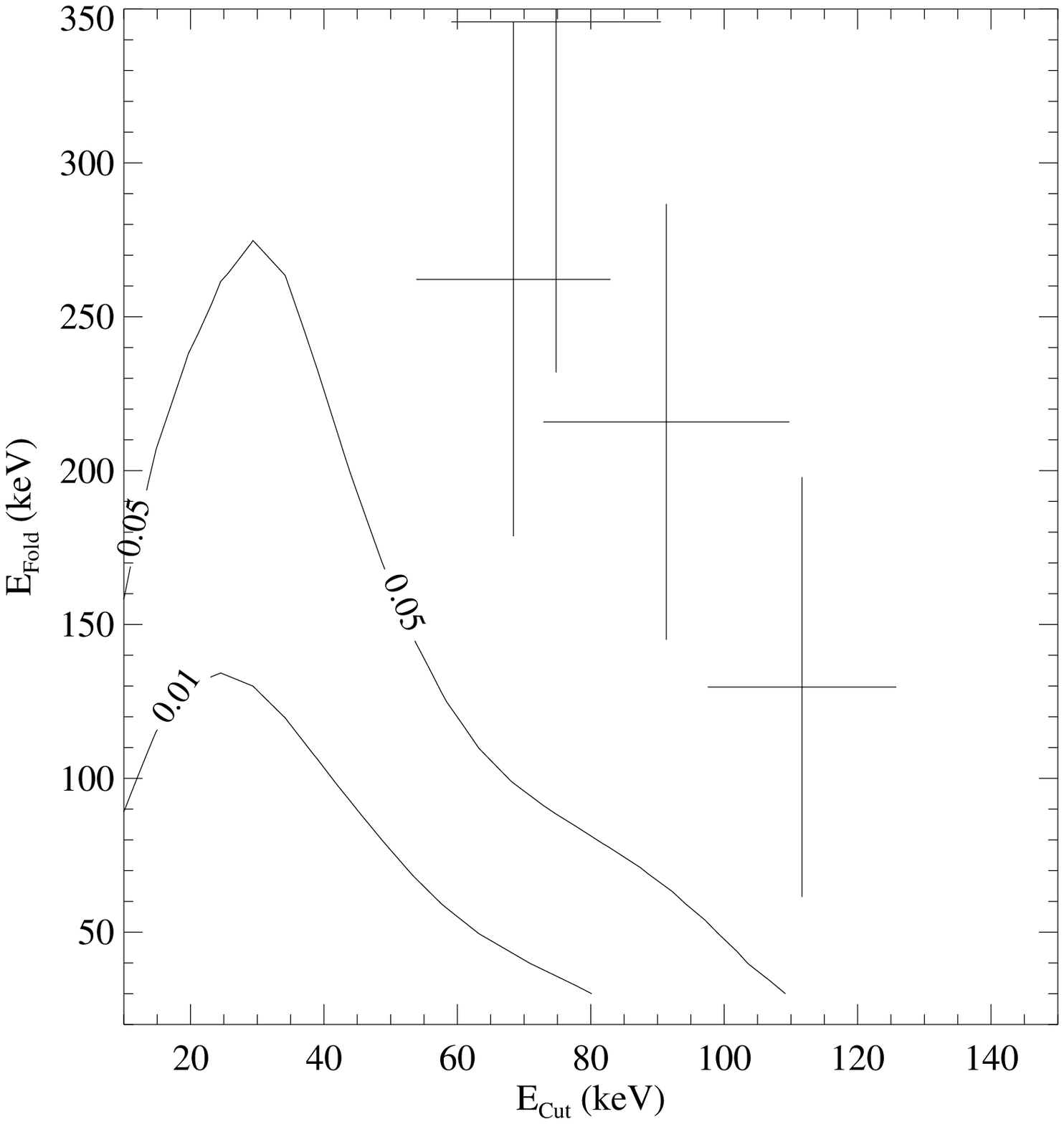}
\caption{\label{fig:cont}
Contours of 5\% and 1\% confidence levels for goodness of fit for the long \HS\ observation taken with \rxte\ when cut-off and folding energies are varied in the fit. The \SX\ cut-off and folding energies with errors are over plotted.
}
\end{figure}


\section{Discussion}\label{sec:discussion}

\subsection{The origin of the flare in X-rays and optical}

The secondary flare is clearly visible at hard X-ray and optical wavelengths, starting around MJD $\wsim 56380$ and ending by MJD $\wsim 56440$ (Figure~\ref{fig:allobs}). Flares in optical and infrared are common during the decays of GBHTs. These flares can be attributed to formation of compact jets \citep[][and references therein]{Kalemci13}, synchrotron radiation from hot accretion flows \citep[][and references therein]{Poutanen13} and enhanced mass accretion from outer disk or the companion as a response to heating from primary \citep{Augusteijn93,Zurita02,Ertan02}.

Being likely a failed outburst, \SX\ did not go through a soft state, and even at the beginning of the decay, the radio emission from the compact jet was present as indicated by the flat to inverted radio spectra from ATCA observations (see Figure~\ref{fig:allobs}). 
While the radio emission is present, whether we could observe infrared emission at the same time depends on the evolution of the break in the jet SED. In MAXI~J1836--194, which also performed a failed outburst, the jet never quenched at radio frequencies, but the break in the spectrum between optically thin and partially self-absorbed (flat / inverted) synchrotron emission moved from the mm regime to the infrared regime as the source made a transition back to the hard state. This indicated an increase in the jet power as the source entered the hard state, resulting in a flare in the infrared and optical emission \citep{Russell13a, RussellTD14}. Similar evolution may be occurring in \SX, whereby the late flare could be a result of break frequency moving towards the infrared. However, the flare in MAXI~J1836--194 peaked within a few days of the transition to the hard state, as did the flare of \GX\ in 2011 \citep{Corbel13}. \cite{Kalemci13} have shown that the infrared flares usually peak about 10-20 days after the source has fully entered the hard state. The late optical and X-ray re-brightening of \SX\ occurred about 100 days after the transition to the hard state, which is therefore substantially delayed compared to these synchrotron jet flares. Unfortunately we do not have optical coverage within days of the transition to the hard state, so we do not know if there was an additional optical/infrared flare at that time. The hot accretion  flow is also an unlikely explanation for the flare, as throughout the decay that lasts for more than 100 days, there is no significant change in the properties of hard X-rays, and it is difficult to explain why the synchrotron emission suddenly turns on 80 days after the start of the decay.

Independent of the emission mechanism, a late flare could be caused by enhanced mass accretion in response to a previous heating event, which could in principal boost all accretion components at late times (disc emission, hot flow or jet). The event that triggered the enhanced mass accretion for \SX\ could be the initial outburst that peaked around MJD~56190, or it could be the increase in the hard X-rays observed between MJD~56260 and MJD~56270 which likely correspond to a transition from an intermediate state to the hard state \citep{Curran14}.  Being observed in both X-rays and optical also support that the origin of this flare is enhanced mass accretion. For such a scenario, one would expect the X-rays to lag the optical flare \citep{Ertan02}, however, the optical coverage is insufficient to test this prediction. 

Such simultaneous X-ray/OIR flares are observed in other sources. \ST\ is such a case in which a late flare occurred $\wsim$55 days after the transition to the hard state. \cite{Russell12} measured the SED optical to X-ray spectral index during the flare to be $-1.0\pm0.3$ which is consistent with synchrotron emission from a compact jet.  A detailed investigation of SMARTS light curves in the $H$ and $I$ bands show that apart from the large flare with simultaneous enhancement in the OIR and X-rays, there is an earlier, smaller flare observed close to the first detection of the compact core with the VLBA \citep{Chun13}. \cite{Kalemci13} interpreted this smaller flare as corresponding to the formation of compact jet, and large simultaneous X-ray/OIR flare as caused by enhanced mass accretion.

Similar behavior is observed during the decay of \GX\ in its 2007 outburst. The OIR flare that started on MJD~ 54241 is interpreted as formation of the compact jet in \cite{Kalemci13}. The OIR flare, instead of decaying within a few tens of days as in other outbursts of \GX, \FU, and \FF\ \citep{Buxton12, Kalemci13} lasted for more than 100 days \citep{Dincer08}. An X-ray flare starts around MJD~52460, that corresponds to a break and a slight increase in the $H$ band flux (See Figure 3 of \citealt{Dincer08}).

Based on these observations, we claim that there are two types of secondary maxima of OIR emission in GBHTs during outburst decays.  During the formation of the compact jets, combination of increase in jet power and/or increase in the frequency of the SED break results in an increase in the OIR emission. These OIR flares are not accompanied by an increase in X-ray flux, and their properties are discussed in detail in \cite{Kalemci13}. The second type of OIR flare is caused by increased mass accretion due to a previous heating event, and these type of flares are accompanied by an increase in X-ray flux. For some sources only one type flare is observed, and for others (like \GX\ in 2007 and \ST) both can be observed during the same outburst.

While the underlying mechanism causing the simultaneous secondary X-ray/OIR flares might be the increased mass accretion for these sources, this does not by itself reveal the emission mechanism. Since the source is in the hard state at this time, a mass increase in the disk going towards the black hole may increase the flux of all the accretion components: inner disk, hot flow and the jet. We might obtain some clues about the emission mechanism by investigating the evolution of X-ray spectra as discussed in the next section.

\subsection{High energy behavior of the source}

Thanks to the wide field of view of the \integral\ ISGRI instrument, and pointed observations with \swift\ XRT, we were able to analyze the broad-band spectral evolution of the source at the beginning of the decay (revolutions 1261-1264), and at the peak of the flare (revolution 1282). We have shown that five revolutions having ISGRI data with good statistical quality require a cut-off in the fit, and we have some evidence (from $\Gamma$) that several of the other probably have a cut-off.

\subsubsection{High energy breaks and jets}

Both of the sources we observed with our \integral\ program, \ST\ and \SX, show presence of a break in the hard X-ray spectrum while the jets are present. This was in contrast to our earlier results from the HEXTE data that indicate disappearance of breaks while the jet turns on.  Suspecting that the disappearance of the break is an artifact of the quality of the data, we decided to check the spectra of \FU\ and \HS\ during their decays to find which combinations of cut-off and folding energies are statistically acceptable. For both \FU\ and \HS\ we found the observations with the best statistics, and multiplied the existing fit with the \emph{highecut}. As reported earlier in \cite{Kalemci05} and \cite{Kalemci06},  the addition of the break into the spectrum is not statistically significant. We then used \emph{steppar} in XSPEC and varied the cut-off and folding energies and recorded the $\chi^{2}$ values. We plotted contours of 5\% and 1\% levels for which the given folding and cut-off values are acceptable. We show the \HS\ obsid 80137-02-01-01 result as an example in Fig.~\ref{fig:cont} because it had $\wsim$6 ks of exposure with \rxte, and had the best statistics for fitting the spectrum.  Even at the 5\% level, a large parameter space of folding and cut-off energies can fit the data. To illustrate this point further, we placed the cut-off and break energy combinations of \SX\ from this work, and all of these combinations would lead to an acceptable fit for the example spectrum of \HS\ that we discuss. For all other observations with \rxte\ (including observations for \FU), the statistics are even worse than the example given, and basically all combinations of cut-off and folding energies are acceptable, except low cut-off ($<$50 keV) and low folding energy ($<$100 keV) combinations. Therefore,  spectra taken with HEXTE on \rxte\ provide insufficient quality to characterize cut-offs during the decay for typical hard X-ray fluxes. Long \integral\ monitoring observations are required to understand the relation between the compact jet formation and hard X-ray behavior.

We can compare the behaviour of the spectral breaks of \SX\ with different sources. For \FU, with the formation of the jet, the folding energy and/or the cut-off energy must have increased. For \HS, no break was necessary in the fit before the presence of compact jets, and this trend continued after the compact jet is observed, so we do not know the evolution, but we can say that even if there is a break, the cut-off and/or folding energies must be high when the compact jet is present. For \SX, we observed an increase in the cut-off energy as the decay progressed during the outburst decay, but we also observed a decrease in the folding energy, resulting in a sharp break in the spectrum. The situation is quite different for \ST\  where we measured a cut-off energy of  $25.3^{+7.3}_{-7.6}$ keV and a folding energy of $236^{+42}_{-32}$ keV \citep{Chun13} indicating a smooth, gradual break at high energies. This observation took place when the compact jet was first observed with the VLBA, but before the large secondary flare. Overall, the behavior of  \SX] and \ST\ for which we have high quality \integral\ data do not support the argument that the high energy breaks disappear as the jets turn on. 

High energy breaks as discussed in this work is a common feature in models invoking both Comptonization and synchrotron as the source of emission. For the model discussed in \cite{Peer12}, the electrons at the jet base characteristically produce a power law with a photon index of around 1.5 and a break at higher energies. The location of the break is determined by the acceleration mechanism whose details are uncertain. While \cite{Peer12} model can roughly reproduce the spectra for both earlier and later revolutions of \SX\ (Asaf Peer, personal communication), it cannot place strong constraints on the physical parameters yet.  

Recently, based on a  broadband observation taken at MJD~56446 at the end of the secondary flare (the last i$^{\prime}$ data in Figure~\ref{fig:allobs}), \cite{Grebenev14} claimed that emission at the time may be dominated by jet emission since a single power-law can fit the entire SED from optical to Gamma-rays. They first take the optical and a single \swift\ BAT flux, and fitted them with a single power-law by modifying the $N_{H}$. Then they added \swift\ XRT spectra taken 10 days earlier (last observation in Table~\ref{xrtonly}), and showed that the single power-law with an $N_{H}$ of $1.205\pm0.026 \, \times 10^{22}$ atoms cm$^{-2}$ found from the optical to BAT fit also fits the XRT data. While this result seems intriguing, we do not think that the quality of the data is sufficient for arriving at strong conclusions about the emission mechanism. We have also analysed the same XRT data using their absorption model, and found that the fit gives $N_{H} = 1.44\pm0.29 \, \times 10^{22}$ atoms cm$^{-2}$ . Therefore errors on both the power-law index and $N_{H}$ are quite large, and it is not surprising that a single power-law passes through all XRT data.

\subsubsection{Comptonization models}

The behavior of hard X-rays can be well explained by Comptonization for \SX. The thermal Comptonization model \emph{compps} fits provide reduced $\chi^{2}$ values close to 1. Finally, all revolutions, except revolution 1262, can be fitted with the \emph{eqpair} model without requiring a hybrid electron plasma. Therefore, except revolution 1262, the high energy behavior of the source is consistent with thermal Comptonization of disk photons. Fig.~\ref{fig:1263fits} show that phenomenological and physical fits all represent the data very well (which is not surprising, see discussion in \citealt{Coppi99}).

Even for revolution 1262 a thermal Comptonization model cannot be ruled out given the very small improvement of $\chi^{2}$ of the hybrid case to that of the thermal case (see Table~\ref{table:joint}). The \emph{compps} still provides an acceptable fit (reduced $\chi^{2}$ of 1.123). We added a hard power-law to \emph{compps} fit (following \citealt{Joinet07}) to test the presence of additional components due to non-thermal Comptonization. The fitted power-law index was $\wsim$1 (not constrained), and the  improvement in the $\chi^{2}$ was marginal (with an $F-test$ chance probability of 0.008). 

While non-thermal electron energy distributions are commonly required in the intermediate and sometimes in the soft states \citep{delSanto08, Gierlinski99, Malzac06}, they are rarely required in the hard states, and all claims of non-thermal electron distributions in the hard state are during bright hard states at the outburst rise \citep{Joinet07, Droulans10, Caballero07, McConnell00}. Thanks to our \integral\ observing program, now we obtained high quality data extending above 300 keV for two sources in the hard state during the decay, \ST\ \citep{Chun13} and \SX, and for both sources fits are consistent with thermal Comptonization. Continuing dedicated \integral\ observations of GBHTs during outburst decays is very important to establish patterns in the high energy behavior, and to compare these patterns with the behavior during outburst rise.

\section{Summary}

We characterized the multiwavelength evolution of \SX\ during the decay of the 2013 outburst using X-ray data from \integral\ ISGRI, JEM-X and \swift\ XRT, optical observations taken at FTS and TUG and radio observations of ATCA. We fit the X-ray spectra with thermal and hybrid Comptonization models as well as phenomenological models. Our main findings can be summarized as follows:

\begin{itemize}

   \item We concluded that the physical origin of a flare observed both in optical and X-rays $\wsim$170 days after the peak of the outburst that lasted for $\wsim$50 days is enhanced mass accretion in response to an earlier heating event. 
   
   \item We showed that a high energy break is needed in the spectra, and for the joint ISGRI-XRT fits, the cut-off energy increased from 75 keV to 112 keV while the folding energy decreased from 350 keV to 130 keV as the decay progresses. 
   
   \item We investigated the claim that high energy cut-offs disappear with the compact jet turning on during outburst decays, and showed that spectra taken with HEXTE on \rxte\ provide insufficient quality to characterize cut-offs during the decay for typical hard X-ray fluxes. Data taken with \integral\ do not support the claim of disappearance of spectral breaks.
   
   \item We found that for the entire decay (including the flare)  the X-ray spectra are consistent with thermal Comptonization, but jet synchrotron origin cannot be ruled out.  

\end{itemize}


\section*{Acknowledgments}
 E.K and T.D. acknowledge T\"UB\.ITAK 1001 Project 111T222, the EU FP7 Initial Training Network Black Hole Universe, ITN 215212.  JAT acknowledges partial support from NASA Astrophysics Data Analysis Program grant NNX11AF84G and \swift\ Guest Observer grant NNX13AJ81G. The Faulkes Telescope South is maintained and operated by Las Cumbres Observatory Global Telescope Network. DMR acknowledges support from a Marie Curie Intra European Fellowship within the 7th European Community Framework Programme (FP7) under contract no. IEF 274805. TMB acknowledges support from INAF PRIN 2012-6. EK thanks J. Chenevez, C. A. Oxborrow and N. Westergaard of DTU Space for JEM-X analysis tips, and A. Pe'er and S. Markoff for valuable discussions on hard X-ray contribution from jets. T.G. acknowledges support from Bilim Akademisi - The Science Academy, Turkey under the BAGEP program.




\end{document}